\documentstyle[12pt,titlepage,epsf]{article}
\def\_{\rule{.3em}{.15ex}} 
\newcommand{\be}{\begin{equation}}
\newcommand{\ee}{\end{equation}}
\newcommand{\bea}{\begin{eqnarray}}
\newcommand{\eea}{\end{eqnarray}}
\newcommand{\f}{\frac}

\newcommand{\al}{\alpha_s}
\newcommand{\bsg}{$b{\to}X_s\gamma $ }
\begin{document}
\begin{titlepage}

 \begin{flushright}
  {\bf ZU-TH 24/96\\
       TUM-HEP-263/96\\       
       MPI/PhT/96-123\\
       hep-ph/9612313\\
       December 1996}
 \end{flushright}

 \begin{center}
  \vspace{0.6in}

\setlength {\baselineskip}{0.3in}
  {\bf \Large Weak Radiative B-Meson Decay Beyond Leading Logarithms}
\vspace{2cm} \\
\setlength {\baselineskip}{0.2in}

{\large  Konstantin Chetyrkin$^{^{1,\dagger}}$, 
         Miko{\l}aj Misiak$^{^{2,\ddagger}}$
         and Manfred M{\"u}nz$^{^{3}}$}\\

\vspace{0.2in}
$^{^{1}}${\it Max-Planck-Institut f{\"u}r Physik, Werner-Heisenberg-Institut,\\
                       D-80805 M{\"u}nchen, Germany}

\vspace{0.2in}
$^{^{2}}${\it Institut f{\"u}r Theoretische Physik der Universit{\"a}t Z{\"u}rich,\\
                        CH-8057 Z{\"u}rich, Switzerland}

\vspace{0.2in}
$^{^{3}}${\it Physik Department, Technische Universit{\"a}t M{\"u}nchen,\\
                         D-85748 Garching, Germany}

\vspace{2cm} 
{\bf Abstract \\} 
\end{center} 
\setlength{\baselineskip}{0.3in} 

	We present our results for three-loop anomalous dimensions
necessary in analyzing $B{\to}X_s\gamma$ decay at the next-to-leading
order in QCD. We combine them with other recently calculated
contributions, obtaining a practically complete next-to-leading order
prediction for the branching ratio ${\cal B}(B{\to}X_s\gamma) = (3.28
\pm 0.33) \times 10^{-4}$. The uncertainty is more than twice smaller
than in the previously available leading order theoretical result. The
Standard Model prediction remains in agreement with the CLEO
measurement at the $2 \sigma$ level.

\vspace{1cm}

\setlength {\baselineskip}{0.2in}
\noindent \underline{\hspace{2in}}\\ 
$^1$ {\footnotesize This work was partially supported by INTAS under 
Contract INTAS-93-0744.}\\
$^2$ {\footnotesize Supported in part by Schweizerischer Nationalfonds
and Polish Commitee for Scientific Research (grant\\
2 P03B 180 09, 1995-97).}\\
$^3$ {\footnotesize Supported in part by Deutsches Bundesministerium
f{\"u}r Bildung und Forschung under contract 06 TM 743.}\\
$^\dagger$ {\footnotesize Permanent address: Institute of Nuclear Research,
Russian Academy of Sciences, Moscow 117312, Russia.}
$^\ddagger$ {\footnotesize Permanent address: Institute of Theoretical
Physics, Warsaw University, Ho\.za 69, 00-681 Warsaw, Poland.}

\end{titlepage} 

\setlength{\baselineskip}{0.3in}

\noindent {\bf 1.} Weak radiative B-meson decay is known to be a very
sensitive probe of new physics \cite{BMMP94,JH93}. Heavy Quark
Effective Theory tells us that inclusive $B$-meson decay rate into
charmless hadrons and the photon is well approximated by the
corresponding partonic decay rate
\be \label{b2B}
\Gamma( B{\to}X_s\gamma) \simeq \Gamma( b{\to}X_s\gamma).
\ee
The accuracy of this approximation is expected to be better than
10\%~\cite{FLS94}.

	The inclusive branching ratio ${\cal B}(b{\to}X_s\gamma)$ was
extracted two years ago from CLEO measurements of weak radiative
$B$-meson decay. It amounts to~\cite{CLEO95}
\be \label{CLEO}
{\cal B}(b{\to}X_s\gamma) = (2.32 \pm 0.57 \pm 0.35) \times 10^{-4}.  
\ee
In the forthcoming few years, much more precise measurements of ${\cal
B}(B{\to}X_s\gamma)$ are expected from the upgraded CLEO detector, as
well as from the B-factories at SLAC and KEK. Acquiring experimental
accuracy of below 10\% is conceivable. Thus, the goal on the
theoretical side is to reach the same accuracy in perturbative
calculations of \bsg decay rate. This requires performing a complete
calculation at the next-to-leading order (NLO) in QCD
\cite{BMMP94,AG92}. All the next-to-leading contributions to \bsg are
collected for the first time in the present letter.\footnote{
\noindent except for the unknown but negligible two-loop matrix
elements of the penguin operators which we denote further by
$P_3$,...,$P_6$.}

	One of the most complex ingredients of the NLO calculation
consists in finding three-loop anomalous dimensions in the effective
theory used for resummation of large logarithms $\; \ln(M_W^2/m_b^2)$.
This is the main new result presented in this letter. However, we have
recalculated all the previously found anomalous dimensions, too.

	The remaining ingredients of the NLO calculation are already
present in the literature \cite{GSW90}--\cite{P96}. We use the
published results for the two-loop matrix elements calculated by
Greub, Hurth and Wyler~\cite{GHW96}, Bremsstrahlung corrections
obtained by Ali, Greub \cite{AG91,AG95} and Pott \cite{P96}, and the
matching conditions found by Adel and Yao \cite{AY94}.

	The following three sections of the present paper are devoted
to presenting the complete NLO formulae for $b{\to}X_s\gamma$. Next,
we analyze them numerically. In the end, we discuss predictions for
the branching ratio ${\cal B}(B{\to}X_s\gamma)$ including a discussion
of the relevant uncertainties. 

\newpage
\noindent {\bf 2.} The analysis of \bsg decay begins with introducing
an effective hamiltonian
\be \label{heff}
H_{eff} = 
-\f{4 G_F}{\sqrt{2}} V^*_{ts} V_{tb} \sum_{i=1}^{8} C_i(\mu) P_i(\mu) 
\ee
\noindent where $V_{ij}$ are elements of the CKM matrix, $P_i(\mu)$
are the relevant operators and $C_i(\mu)$ are the corresponding Wilson
coefficients. The complete set of physical operators necessary for
\bsg decay is the following:
\be \label{ope}
\begin{array}{rl}
P_1 = & (\bar{s}_L \gamma_{\mu} T^a c_L) (\bar{c}_L \gamma^{\mu} T^a b_L) 
\vspace{0.2cm} \\
P_2 = & (\bar{s}_L \gamma_{\mu}     c_L) (\bar{c}_L \gamma^{\mu}     b_L) 
\vspace{0.2cm} \\
P_3 = & (\bar{s}_L \gamma_{\mu}     b_L) \sum_q (\bar{q}\gamma^{\mu}     q)     
\vspace{0.2cm} \\
P_4 = & (\bar{s}_L \gamma_{\mu} T^a b_L) \sum_q (\bar{q}\gamma^{\mu} T^a q)     
\vspace{0.2cm} \\
P_5 = & (\bar{s}_L \gamma_{\mu_1}
                   \gamma_{\mu_2}
                   \gamma_{\mu_3}    b_L)\sum_q (\bar{q} \gamma^{\mu_1} 
                                                         \gamma^{\mu_2}
                                                         \gamma^{\mu_3}     q)     
\vspace{0.2cm} \\
P_6 = & (\bar{s}_L \gamma_{\mu_1}
                   \gamma_{\mu_2}
                   \gamma_{\mu_3} T^a b_L)\sum_q (\bar{q} \gamma^{\mu_1} 
                                                          \gamma^{\mu_2}
                                                          \gamma^{\mu_3} T^a q)     
\vspace{0.2cm} \\
P_7  = &  \f{e}{16 \pi^2} m_b (\bar{s}_L \sigma^{\mu \nu}     b_R) F_{\mu \nu} 
\vspace{0.2cm} \\
P_8  = &  \f{g}{16 \pi^2} m_b (\bar{s}_L \sigma^{\mu \nu} T^a b_R) G_{\mu \nu}^a 
\end{array}
\ee
where $T^a$ stand for $SU(3)_{\rm color}$ generators. The small CKM
matrix element $V_{ub}$ as well as the $s$-quark mass are neglected in
the present paper.

	Our basis of four-quark operators is somewhat different than
the standard one used in refs.~\cite{BMMP94,GSW90,GW79}, although the
two bases are physically equivalent. We would like to stress that in
our basis, Dirac traces containing $\gamma_5$ do not arise in
effective theory calculations performed at the leading order in the
Fermi coupling $G_F$ (but to all orders in QCD). This allows us to
consistently use fully anticommuting $\gamma_5$ in dimensional
regularization, which greatly simplifies multiloop calculations. Our
choice of color structures is dictated only by convenience in computer
algebra applications.

	Throughout the paper, we use the $\overline{MS}$ scheme with
fully anticommuting $\gamma_5$. Such a scheme is not uniquely defined
until one chooses a specific form for the so-called evanescent
operators which algebraically vanish in four dimensions
\cite{DG91,HN95}. In our three-loop calculation, eight evanescent
operators were necessary. We list them explicitly in Appendix A.

	Resummation of large logarithms $\ln(M_W^2/m_b^2)$ is
achieved by evolving the coefficients $C_i(\mu)$ from $\mu = M_W$ to
$\mu = \mu_b \simeq m_b$ according to the Renormalization Group
Equations (RGE). Instead of the original coefficients $C_i(\mu)$, it
is convenient to use certain linear combinations of them, called the
``effective coefficients'' \cite{BMMP94,GHW96}
\be
C_i^{eff}(\mu) = \left\{ \begin{array}{cc}
C_i(\mu) & \mbox{ for $i = 1, ..., 6$} \\ 
C_7(\mu) + \sum_{i=1}^6 y_i C_i(\mu) & \mbox{ for $i = 7$} \\
C_8(\mu) + \sum_{i=1}^6 z_i C_i(\mu) & \mbox{ for $i = 8$.}
\end{array} \right.
\ee
The numbers $y_i$ and $z_i$ are defined so that the leading-order $b
\to s \gamma$ and $b \to s\;gluon$ matrix elements of the effective
hamiltonian are proportional to the leading-order terms in $C_7^{eff}$
and $C_8^{eff}$, respectively \cite{BMMP94}. In the $\overline{MS}$
scheme with fully anticommuting $\gamma_5$, we have \linebreak[4] $y =
(0, 0, -\f{1}{3}, -\f{4}{9}, -\f{20}{3}, -\f{80}{9})$ and $z = (0, 0,
1, -\f{1}{6}, 20, -\f{10}{3})$.\footnote{ These numbers are different
than in section 4 of ref.~\cite{BMMP94} because we use a different
basis of four-quark operators here.} The leading-order contributions
to the effective coefficients $C_i^{eff}(\mu)$ are regularization- and
renormalization-scheme independent, which would not be true for the
original coefficients $C_7(\mu)$ and $C_8(\mu)$.

	The effective coefficients evolve according to their RGE
\be \label{RGE}
\mu \f{d}{d \mu} C_i^{eff}(\mu) = C_j^{eff}(\mu) \gamma^{eff}_{ji}(\mu)
\ee
driven by the anomalous dimension matrix
$\hat{\gamma}^{eff}(\mu)$. One expands this matrix perturbatively as
follows:
\be 
\hat{\gamma}^{eff}(\mu) = \f{\al (\mu)}{4 \pi} \hat{\gamma}^{(0)eff}  
                        + \f{\al^2(\mu)}{(4 \pi)^2} \hat{\gamma}^{(1)eff} + ...  
\ee
The matrix $\hat{\gamma}^{(0)eff}$ is renormalization-scheme
independent, while $\hat{\gamma}^{(1)eff}$ is not. In the
$\overline{MS}$ scheme with fully anticommuting $\gamma_5$ (and with
our choice of evanescent operators) we obtain
\be \label{gamma0}
\hat{\gamma}^{(0)eff} = \left[
\begin{array}{cccccccc}
\vspace{0.2cm}
-4 & \f{8}{3} &       0     &   -\f{2}{9} &      0    &     0     & -\f{208}{243} &  \f{173}{162} \\ 
\vspace{0.2cm}
12 &     0    &       0     &    \f{4}{3} &      0    &     0     &   \f{416}{81} &    \f{70}{27} \\ 
\vspace{0.2cm}
 0 &     0    &       0     &  -\f{52}{3} &      0    &     2     &  -\f{176}{81} &    \f{14}{27} \\ 
\vspace{0.2cm}
 0 &     0    &  -\f{40}{9} & -\f{100}{9} &  \f{4}{9} &  \f{5}{6} & -\f{152}{243} & -\f{587}{162} \\ 
\vspace{0.2cm}
 0 &     0    &       0     & -\f{256}{3} &      0    &    20     & -\f{6272}{81} &  \f{6596}{27} \\ 
\vspace{0.2cm}
 0 &     0    & -\f{256}{9} &   \f{56}{9} & \f{40}{9} & -\f{2}{3} & \f{4624}{243} &  \f{4772}{81} \\ 
\vspace{0.2cm}
 0 &     0    &       0     &       0     &      0    &     0     &     \f{32}{3} &        0      \\ 
\vspace{0.2cm}
 0 &     0    &       0     &       0     &      0    &     0     &    -\f{32}{9} &     \f{28}{3} \\
\end{array} \right] \ee
and
\be \label{gamma1}
\hat{\gamma}^{(1)eff} = \left[
\begin{array}{cccccccc}
\vspace{0.2cm}
-\f{355}{9} & -\f{502}{27} &  -\f{1412}{243} &  -\f{1369}{243} &    \f{134}{243} &   -\f{35}{162} &     -\f{818}{243} &     \f{3779}{324} \\ 
\vspace{0.2cm}
 -\f{35}{3} &   -\f{28}{3} &    -\f{416}{81} &    \f{1280}{81} &      \f{56}{81} &     \f{35}{27} &       \f{508}{81} &     \f{1841}{108} \\ 
\vspace{0.2cm}
     0      &        0     &   -\f{4468}{81} &  -\f{31469}{81} &     \f{400}{81} &  \f{3373}{108} &    \f{22348}{243} &     \f{10178}{81} \\ 
\vspace{0.2cm}
     0      &        0     &  -\f{8158}{243} & -\f{59399}{243} &    \f{269}{486} & \f{12899}{648} &   -\f{17584}{243} &  -\f{172471}{648} \\ 
\vspace{0.2cm}
     0      &        0     & -\f{251680}{81} & -\f{128648}{81} &   \f{23836}{81} &   \f{6106}{27} &  \f{1183696}{729} &  \f{2901296}{243} \\ 
\vspace{0.2cm}
     0      &        0     &  \f{58640}{243} & -\f{26348}{243} & -\f{14324}{243} & -\f{2551}{162} & \f{2480344}{2187} & -\f{3296257}{729} \\ 
\vspace{0.2cm}
     0      &        0     &         0       &         0       &         0       &        0       &      \f{4688}{27} &          0        \\  
\vspace{0.2cm}
     0      &        0     &         0       &         0       &         0       &        0       &     -\f{2192}{81} &     \f{4063}{27} \\
\end{array} \right]. \ee
The matrix $\hat{\gamma}^{(1)eff}$ is the main new result of this
paper. Its evaluation required performing three-loop renormalization
of the effective theory (\ref{heff}). Details of this calculation will
be given in a forthcoming publication \cite{CMM96}.

	We expand the coefficients in powers of $\al$ as follows:
\be \label{C.expanded}
C^{eff}_i(\mu) = C^{(0)eff}_i(\mu) + \f{\al(\mu)}{4 \pi} C^{(1)eff}_i(\mu) + ...
\ee
At $\mu = M_W$, the values of the coefficients are found by matching
the effective theory amplitudes with the full Standard Model ones. The
well-known leading-order results read \cite{BMMP94,GSW90,IL81}
\be
C_i^{(0)eff}(M_W) = C_i^{(0)}(M_W) = \left\{ \begin{array}{cc}
0 & \mbox{ for $i = 1,3,4,5,6$} \\
1 & \mbox{ for $i = 2$} \\
\vspace{0.2cm} 
\f{3 x^3 - 2 x^2}{4 (x-1)^4} \ln x + \f{-8 x^3 - 5 x^2 + 7 x}{24 (x-1)^3}
                                                  & \mbox{ for $i = 7$} \\
\f{- 3 x^2}{4 (x-1)^4} \ln x + \f{-x^3 + 5 x^2 + 2 x}{8 (x-1)^3}
                                                  & \mbox{ for $i = 8$}, \\
\end{array} \right.
\ee
where 
\be
x = \f{m_{t,\overline{MS}}^2(M_W)}{M_W^2} = \f{m_{t,pole}^2}{M_W^2} 
\left( \f{\al(M_W)}{\al(m_t)} \right)^{\f{24}{23}} 
\left( 1 - \f{8}{3} \f{\al(m_t)}{\pi} \right).
\ee

	The next-to-leading contributions to the four-quark operator
coefficients at $\mu = M_W$ can be found either by transforming the
results of ref.~\cite{BJLW92} to our operator basis, or by a direct
computation. We find
\be
C_i^{(1)eff}(M_W) = C_i^{(1)}(M_W) = \left\{ \begin{array}{cc}
15 & \mbox{ for $i = 1$} \\
0 & \mbox{ for $i = 2,3,5,6$} \\
E(x) - \f{2}{3} & \mbox{ for $i = 4$}, 
\end{array} \right.
\ee
where
\be
E(x) = \f{x (18 -11
x - x^2)}{12 (1-x)^3} + \f{x^2 (15 - 16 x + 4 x^2)}{6 (1-x)^4} \ln
x-\f{2}{3} \ln x.
\ee

	The next-to-leading contributions to $C_7^{eff}(M_W)$ and
$C_8^{eff}(M_W)$ can be extracted from ref.~\cite{AY94} according to
the following equations:
\be
C_7^{(1)eff}(M_W) = -8/3 \f{\pi}{4 \al} \bar{C}^{(1)}_{O_{52}} +
\Delta_7 - \f{4}{9} \left( E(x) - \f{2}{3} \right),
\ee
\be
C_8^{(1)eff}(M_W) = 8 \f{\pi}{4 \al} \bar{C}^{(1)}_{O_{51}} +
\Delta_8 - \f{1}{6} \left( E(x) - \f{2}{3} \right),
\ee
where the quantities $\bar{C}^{(1)}_{O_{51}}$ and
$\bar{C}^{(1)}_{O_{52}}$ are given in eqn.~(5) of ref.~\cite{AY94} with
$\mu=M_W$. The terms denoted by $\Delta_7$ and $\Delta_8$ stand for
shifts one needs to make when passing from the so-called $R^*$
renormalization scheme used in ref.~\cite{AY94} to the $\overline{MS}$
scheme used here
\be 
\Delta_i = \f{4 \pi}{\al} 
\left[ C_i^{(0)}(M_W)( x \to x - \f{2 \al}{\pi} x \ln x ) 
    -  C_i^{(0)}(M_W)( x ) \right] \hspace{1cm} \mbox{ for $i = 7,8$}.
\ee
Only the leading (zeroth-order) term in $\al$ needs to be retained in
the above equation.

	The resulting explicit expressions for $C_7^{eff}(M_W)$ and
$C_8^{eff}(M_W)$ in the $\overline{MS}$ scheme read
\bea
C_7^{(1)eff}(M_W) &=& \f{-16 x^4 -122 x^3 + 80 x^2 -  8 x}{9 (x-1)^4} {\rm Li}_2 \left( 1 - \f{1}{x} \right)
                  +\f{6 x^4 + 46 x^3 - 28 x^2}{3 (x-1)^5} \ln^2 x 
\nonumber \\ &&
                  +\f{-102 x^5 - 588 x^4 - 2262 x^3 + 3244 x^2 - 1364 x + 208}{81 (x-1)^5} \ln x
\nonumber \\ &&
                  +\f{1646 x^4 + 12205 x^3 - 10740 x^2 + 2509 x - 436}{486 (x-1)^4} 
\vspace{0.2cm} \\
C_8^{(1)eff}(M_W) &=& \f{-4 x^4 +40 x^3 + 41 x^2 + x}{6 (x-1)^4} {\rm Li}_2 \left( 1 - \f{1}{x} \right)
                  +\f{ -17 x^3 - 31 x^2}{2 (x-1)^5} \ln^2 x 
\nonumber \\ &&
                  +\f{ -210 x^5 + 1086 x^4 +4893 x^3 + 2857 x^2 - 1994 x +280}{216 (x-1)^5} \ln x
\nonumber \\ &&
                  +\f{737 x^4 -14102 x^3 - 28209 x^2 + 610 x - 508}{1296 (x-1)^4}
\eea

	Having specified the initial conditions at $\mu = M_W$, we are
ready to write the solution to the Renormalization Group Equations
(\ref{RGE}) for $C^{eff}_7(\mu_b)$
\be      \label{c7eff0}
C^{(0)eff}_7(\mu_b) = \eta^{\f{16}{23}} C^{(0)}_7(M_W) +
\f{8}{3} \left( \eta^{\f{14}{23}} - \eta^{\f{16}{23}}
\right) C^{(0)}_8(M_W)  + \sum_{i=1}^8 h_i \eta^{a_i}, 
\ee
\bea     \label{c7eff1}
C^{(1)eff}_7(\mu_b) &=& 
\eta^{\f{39}{23}} C^{(1)eff}_7(M_W) + \f{8}{3} \left( \eta^{\f{37}{23}} - \eta^{\f{39}{23}} \right) C^{(1)eff}_8(M_W) 
\nonumber \\ &&
+\left( \f{297664}{14283} \eta^{\f{16}{23}}-\f{7164416}{357075} \eta^{\f{14}{23}} 
       +\f{256868}{14283} \eta^{\f{37}{23}} -\f{6698884}{357075} \eta^{\f{39}{23}} \right) C_8^{(0)}(M_W) 
\nonumber \\ &&
+\f{37208}{4761} \left( \eta^{\f{39}{23}} - \eta^{\f{16}{23}} \right) C_7^{(0)}(M_W) 
+ \sum_{i=1}^8 (e_i \eta E(x) + f_i + g_i \eta) \eta^{a_i},
\eea
where $\eta = \al(M_W)/\al(\mu_b)$, and the numbers $a_i$--$h_i$ are
as follows
\bea  
\begin{array}{ccccccccc}
\vspace{0.2cm}
a_i = ( &    \f{14}{23},     &      \f{16}{23},   & \f{6}{23},& -\f{12}{23}, 
        &      0.4086,       &       -0.4230,     & -0.8994,  &   0.1456   )\\
\vspace{0.2cm}
e_i = ( &\f{4661194}{816831},&   -\f{8516}{2217}, &     0,    &     0, 
        &     -1.9043,       &       -0.1008,     &  0.1216,  &  0.0183    )\\
\vspace{0.2cm}
f_i = ( &    -17.3023,       &        8.5027,     &  4.5508,  &  0.7519,
        &      2.0040,       &        0.7476,     & -0.5385,  &  0.0914    )\\
\vspace{0.2cm}
g_i = ( &     14.8088,       &      -10.8090,     & -0.8740,  &  0.4218, 
        &     -2.9347,       &        0.3971,     &  0.1600,  &  0.0225    )\\
\vspace{0.2cm}
h_i = ( &\f{626126}{272277}, & -\f{56281}{51730}, & -\f{3}{7},& -\f{1}{14},
        &     -0.6494,       &       -0.0380,     & -0.0186,  & -0.0057    ).
\end{array}
&& \nonumber \eea

	As far as the other Wilson coefficients are concerned, only
the leading-order contributions to them are necessary in the complete
NLO analysis of $b{\to}X_s\gamma$.  We obtain\footnote{ Analogous expressions in
ref.~\cite{BMMP94} are somewhat different, because we use a different
basis of four-quark operators here.}
\bea
\begin{array}{ccc}
\vspace{0.2cm}
C^{(0)}_1(\mu_b)=& \hspace{-0.4cm} 
		 -\eta^{-\f{12}{23}} + \eta^{\f{6}{23}},&\\
\vspace{0.2cm}
C^{(0)}_2(\mu_b)=& \hspace{-0.1cm}  
                  \f{1}{3} \eta^{-\f{12}{23}} + \f{2}{3} \eta^{\f{6}{23}},&\\
\vspace{0.2cm}
C^{(0)}_3(\mu_b)=& \hspace{-0.1cm}  
                -\f{1}{27} \eta^{-\f{12}{23}} + \f{2}{63} \eta^{\f{6}{23}}&  
\hspace{-0.4cm} - 0.0659 \eta^{0.4086}   + 0.0595 \eta^{-0.4230} 
                - 0.0218 \eta^{-0.8994}  + 0.0335 \eta^{0.1456},\\ 
\vspace{0.2cm}
C^{(0)}_4(\mu_b)=& \hspace{-0.4cm} 
                 \f{1}{9} \eta^{-\f{12}{23}} + \f{1}{21} \eta^{\f{6}{23}}&  
\hspace{-0.4cm} + 0.0237 \eta^{0.4086}   - 0.0173 \eta^{-0.4230} 
                - 0.1336 \eta^{-0.8994}  - 0.0316 \eta^{0.1456},\\
\vspace{0.2cm}
C^{(0)}_5(\mu_b)=& \hspace{-0.4cm} 
                 \f{1}{108} \eta^{-\f{12}{23}} - \f{1}{126} \eta^{\f{6}{23}}& 
\hspace{-0.4cm} + 0.0094 \eta^{0.4086}   - 0.0100 \eta^{-0.4230} 
                + 0.0010 \eta^{-0.8994}  - 0.0017 \eta^{0.1456},\\
\vspace{0.2cm}
C^{(0)}_6(\mu_b)=& \hspace{-0.4cm} 
                -\f{1}{36} \eta^{-\f{12}{23}} - \f{1}{84} \eta^{\f{6}{23}}&  
\hspace{-0.4cm} + 0.0108 \eta^{0.4086}   + 0.0163 \eta^{-0.4230} 
                + 0.0103 \eta^{-0.8994}  + 0.0023 \eta^{0.1456},
\end{array} \nonumber 
\eea
\vspace*{-1.8cm}\\ 
\bea
C^{(0)eff}_8(\mu_b) = \left( C^{(0)}_8(M_W)  + \f{313063}{363036} \right) 
\eta^{\f{14}{23}} \hspace{10cm}
\nonumber \end{eqnarray} \vspace*{-1.3cm}
\begin{eqnarray} \hspace{5.3cm} 
              -0.9135 \eta^{0.4086}   +  0.0873 \eta^{-0.4230}       
              -0.0571 \eta^{-0.8994}  +  0.0209 \eta^{0.1456}. 
\eea 

	In our numerical analysis, the values of $\al(\mu)$ in all the
above formulae are calculated with use of the NLO expression for the
strong coupling constant
\be \label{alphaNLL}
\al(\mu) = \f{\al(M_Z)}{v(\mu)} \left[1 - \f{\beta_1}{\beta_0} 
           \f{\al(M_Z)}{4 \pi}    \f{\ln v(\mu)}{v(\mu)} \right],
\ee
where 
\be \label{v(mu)}
v(\mu) = 1 - \beta_0 \f{\al(M_Z)}{2 \pi} \ln \left( \f{M_Z}{\mu} \right),
\ee
$\beta_0 = \f{23}{3}$ and $\beta_1 = \f{116}{3}$.

	As easily can be seen from eqns.
(\ref{C.expanded})--(\ref{v(mu)}), the Wilson coefficients at $\mu =
\mu_b$ depend on only five parameters: $M_Z$, $M_W$, $\al(M_Z)$,
$m_{t,pole}$ and $\mu_b$. In our numerical analysis, the first two of
them are fixed to $M_Z=91.187$~GeV and $M_W=80.33$~GeV \cite{PDG96}.
For the remaining three, we take
\vspace{-0.5cm}
\bea \label{alpha.and.mt}
\al(M_Z) &=& 0.118 \pm 0.003 \hspace{1cm} \mbox{\cite{S96}},\\
m_{t,pole} &=& 175 \pm 6 \; {\rm GeV} \hspace{1cm} \mbox{\cite{T96}}
\eea
\ \vspace{-1cm}\\
and \vspace{-0.5cm}
\be \label{mu.range}
\mu_b \in [ 2.5\;{\rm GeV}, 10\;{\rm GeV}].
\ee
For $\mu_b = 5$ GeV and central values of the other parameters, we
find $\; \al(\mu_b) = 0.212$,
\bea
\overline{C}^{(0)eff}(\mu_b) &=& \begin{array}{cccccccc}
(-0.480, & 1.023, & -0.0045, & -0.0640, & 0.0004, & 0.0009, & -0.312, & -0.148),
\end{array} \nonumber \\
C^{(1)eff}_7(\mu_b) &=& -0.995 + 1.443 \;\;=\;\; 0.448.
\eea
In the intermediate step of the latter equation, we have shown
relative importance of two contributions to $C^{(1)eff}_7(\mu_b)$. The
first of them originates from terms proportional to
$C^{(1)eff}_7(M_W)$ and $C^{(1)eff}_8(M_W)$ in eqn.~(\ref{c7eff1}),
i.e. it is due to the NLO matching. The second of them comes from the
remaining terms, i.e. it is driven by the next-to-leading anomalous
dimensions. Quite accidentally, the two terms tend to cancel each
other. In effect, the total influence of $C^{(1)eff}_7(\mu_b)$ on the
\bsg decay rate does not exceed 6\%. (Explicit expressions for the
decay rate are given below). However, if the relative sign between the
two contributions was positive, then the decay rate would be affected
by almost 30\%. This observation illustrates the importance of
explicit calculation of both the NLO matching conditions and the NLO
anomalous dimensions.\vspace{0.5cm}

\noindent {\bf 3.} At this point, we are ready to use the calculated
coefficients at $\mu = \mu_b \simeq m_b$ as an input in the NLO
expressions for \bsg decay rate given by Greub, Hurth and
Wyler~\cite{GHW96}. However, we make an explicit lower cut on the
photon energy in the Bremsstrahlung correction
\be
E_{\gamma} > ( 1 - \delta ) E_{\gamma}^{max} \equiv ( 1 - \delta ) \f{m_b}{2}.
\ee

	We write the decay rate as follows:
\bea \label{rate}
\Gamma[ b{\to}X_s\gamma]^{
E_{\gamma} > (1-\delta) E_{\gamma}^{max}} &=&
\Gamma[ b \to s \gamma] \;\; + \;\; \Gamma[ b \to s \gamma \; gluon]^{
E_{\gamma} > (1-\delta) E_{\gamma}^{max}} \;\;= 
\vspace{0.2cm} \nonumber \\
&=& \f{G_F^2 \alpha_{em}}{32 \pi^4} |V^*_{ts} V_{tb}|^2
m_{b,pole}^3 m_{b,\overline{MS}}^2(m_b) \left( |D|^2 + A \right),
\eea
where 
\be \label{Dvirt}
D = C_7^{(0)eff}(\mu_b) + \f{\al(\mu_b)}{4 \pi} \left\{ 
C_7^{(1)eff}(\mu_b) + \sum_{i=1}^8 C_i^{(0)eff}(\mu_b) \left[ 
r_i + \gamma_{i7}^{(0)eff} \ln \f{m_b}{\mu_b} 
\right] \right\}
\ee
and
\be \label{Abrem}
A = \left( e^{ -\al(\mu_b) \ln \delta (7 + 2 \ln \delta)/ 3 \pi} - 1 \right)
|C_7^{(0)eff}(\mu_b)|^2  \; + \f{\al(\mu_b)}{\pi} 
\sum_{ \stackrel{i,j=1}{i \leq j}}^8
         C_i^{(0)eff}(\mu_b) C_j^{(0)eff}(\mu_b) f_{ij}(\delta).
\ee
The contribution denoted by $|D|^2$ is independent of the cutoff
parameter $\delta$. By differentiating $D$ with respect to $\mu_b$ and
using the RGE (\ref{RGE}), one can easily find out that the leading
$\mu_b$-dependence cancels out in $D$. When calculating $|D|^2$ in our
numerical analysis, we consistently set the ${\cal O}(\al^2)$
term to zero.

	The contribution denoted by $A$ vanishes in the limit
$\al{\to}0$. The first term in $A$ contains the exponentiated
(infrared) logarithms of $\delta$ which remain after the IR
divergences cancel between the virtual and Bremsstrahlung corrections
to $b{\to}X_s\gamma$.

	The terms denoted by $r_i$ in $D$ depend on what convention is
chosen for additive constants in the functions $f_{i7}(\delta)$ and
$f_{78}(\delta)$. We fix this convention by requiring that all
$f_{ij}(\delta)$ vanish in the formal limit $\delta \to 0$.

	In order to complete the full presentation of \bsg results, it
remains to give explicitly the constants $r_i$ and the functions
$f_{ij}(\delta)$. 

	The constants $r_8$ and $r_2$ are exactly as given in
ref.~\cite{GHW96}
\bea
\hspace{1.2cm}
r_8 &=& -\f{4}{27} ( -33 + 2 \pi^2 - 6 i \pi ), \\
r_2  &=& \frac{2}{243} \, \left\{ -833 + 144 \pi^2 z^{3/2} \right. 
\nonumber \\
&& \hspace{1.2cm}
+ \left[ 1728 -180 \pi^2 -1296 \zeta (3) + (1296 - 324 \pi^2) L +
108 L^2 + 36 L^3 \right] \, z 
\nonumber \\
&& \hspace{1.2cm}
+ \left[ 648 + 72 \pi^2 + (432 - 216 \pi^2) L + 36 L^3 \right] \, z^2
\nonumber \\
&& \hspace{1.2cm}        \left.                 +
\left[ -54 - 84 \pi^2 + 1092 L - 756 L^2 \right] \, z^3 \, \right\}
\nonumber \\
&& + \frac{16 \pi i}{81} \, \left\{ -5 +
\left[ 45 -3 \pi^2 + 9 L + 9 L^2 \right] \, z \right. 
\nonumber \\
&& \hspace{1.2cm} + \left[ -3 \pi^2 + 9 L^2 \right] \, z^2 
\nonumber \\
&& \hspace{1.2cm} + \left.
\left[ 28 - 12 L  \right] \, z^3 \, \right\} 
\hspace{0.3cm} + \hspace{0.3cm} {\cal O}(z^4 L^4),
\eea
where $z = m_c^2/m_b^2$ and $L = \ln z$. In our operator basis
(\ref{ope}), the constant $r_1$ does not vanish, but is proportional
to $r_2$
\be
r_1 = -\f{1}{6} r_2.
\ee
The constant $r_7$ turns out to be\footnote{
It differs from the one given in ref.~\cite{GHW96} where a formal
limit $ \delta \to 1$ was taken and the contribution due to
$f_{77}(1)$ was absorbed into $r_7$.}
\be
r_7 = -\f{10}{3} -\f{8}{9} \pi^2.
\ee
The quantities $r_3$, ..., $r_6$ originate from two-loop matrix
elements of the penguin operators $P_3$, ..., $P_6$. They remain the
only still unknown elements of the formally complete NLO analysis of
$b{\to}X_s\gamma$. However, the Wilson coefficients $C_i$ are very
small for $i = 3,\;...,\;6$. In consequence, setting $r_3$, ..., $r_6$
to zero (which we do in our numerical analysis) has a negligible
effect, i.e. it can potentially affect the decay rate by only about
1\%.

	Let us now turn to the functions $f_{ij}(\delta)$. One can
easily obtain $f_{77}(\delta)$ from eqns. (23) and (24) of
ref.~\cite{AG95}. (These equations were also the source for our $r_7$
and the exponent in eqn.~(\ref{Abrem}) above.)
\be
f_{77}(\delta) = \f{10}{3} \delta + \f{1}{3} \delta^2 - \f{2}{9} \delta^3
+ \f{1}{3} \delta ( \delta - 4 ) \ln \delta.
\ee
The function $f_{88}(\delta)$ can be found by integrating eqn.~(18) of
the same paper \cite{AG95} 
\bea 
f_{88}(\delta) &=& \f{1}{27} \left\{ - 2 \ln \f{m_b}{m_s} 
                     \left[ \delta^2 + 2 \delta + 4 \ln(1-\delta) \right] 
\right. \nonumber \\ && \left.
+4{\rm Li}_2(1-\delta) -\f{2\pi^2}{3} -\delta(2+\delta)\ln\delta + 8\ln(1-\delta)
-\f{2}{3} \delta^3 + 3 \delta^2 + 7 \delta \right\}. \label{f88}
\eea
The logarithm of $m_s$ in the above equation is the only point in our
analysis, at which the $s$-quark mass cannot be neglected. Collinear
divergences which arise in the massless $s$-quark limit have been
discussed in ref.~\cite{KLP95}. The properly resumed photon spectrum
was there found to be finite in the $m_s \to 0$ limit, and suppressed
with respect to the case when $\ln(m_b/m_s)$ in $f_{88}(\delta)$ is
used. Even with $m_s$ as small as $0.1$ GeV, the term proportional to
$\ln(m_b/m_s)$ is negligible, i.e. it affects the decay rate by less
than 1\%. Therefore, for simplicity, we will just use the naive
expression (\ref{f88}) with $m_b/m_s = 50$ in our numerical analysis.

	The functions $f_{22}(\delta)$, $f_{27}(\delta)$,
$f_{28}(\delta)$ and $f_{78}(\delta)$ can be found from appendix B of
ref.~\cite{GHW96}. After performing some of the phase-space
integrations, one finds
\bea
f_{22}(\delta) &=& \f{16 z}{27} \left[ 
\delta \int_0^{(1-\delta)/z} dt \; (1-zt) \left| \f{G(t)}{t} + \f{1}{2} \right|^2 
\;+\;
\int_{(1-\delta)/z}^{1/z} dt \; (1-zt)^2  \left| \f{G(t)}{t} + \f{1}{2} \right|^2 
\right], 
\nonumber \\
f_{27}(\delta) &=& -\f{8 z^2}{9} \left[ 
\delta \int_0^{(1-\delta)/z} dt\;    {\rm Re} \left( G(t) + \f{t}{2} \right) \;+\;
\int_{(1-\delta)/z}^{1/z} dt\;(1-zt) {\rm Re} \left( G(t) + \f{t}{2} \right) \right],
\nonumber \\
f_{28}(\delta) &=& -\f{1}{3} \; f_{27}(\delta),
\nonumber \\
f_{78}(\delta) &=& \f{8}{9} \left[ {\rm Li}_2(1-\delta) - \f{\pi^2}{6} - \delta 
\ln \delta + \f{9}{4} \delta - \f{1}{4} \delta^2 + \f{1}{12} \delta^3 \right],
\eea
where, as before, $z = m_c^2/m_b^2$ and
\be
G(t) = \left\{ \begin{array}{cc} 
- 2 \arctan^2 \sqrt{ t/(4-t)}, & \mbox{ for $t < 4$} \vspace{0.2cm} \\
-\pi^2/2 + 2 \ln^2[(\sqrt{t} + \sqrt{t-4})/2] 
- 2 i \pi \ln[(\sqrt{t} + \sqrt{t-4})/2], & \mbox{ for $t \geq 4$}. 
\end{array} \right.
\ee

	In our operator basis (\ref{ope}), the functions
$f_{1j}(\delta)$ do not vanish, but are proportional to the functions
$f_{2j}(\delta)$
\bea
\begin{array}{cccc}
f_{11}(\delta) = \f{1}{36} f_{22}(\delta), \hspace{0.7cm}&
f_{12}(\delta) = -\f{1}{3} f_{22}(\delta), \hspace{0.7cm}&
f_{17}(\delta) = -\f{1}{6} f_{27}(\delta), \hspace{0.7cm}&
f_{18}(\delta) = -\f{1}{6} f_{28}(\delta).
\end{array} \nonumber \eea

	The only functions $f_{ij}(\delta)$ which have not yet been
given explicitly are the ones with at least one index corresponding to
the penguin operators $P_3$, ..., $P_6$. As checked in
ref.~\cite{P96}, these functions cannot affect the decay rate by more
than 2\%. Therefore, we neglect them in our numerical analysis.
However, for completeness, a formula from which they can be obtained
is given in our Appendix B. \vspace{0.5cm}

\noindent {\bf 4.} The decay rate given in eqn.~(\ref{rate}) suffers
from large uncertainties due to $m_{b,pole}^5$ and the CKM angles. One
can get rid of them by normalizing \bsg decay rate to the semileptonic
decay rate of the b-quark
\be
\Gamma[ b \to X_c e \bar{\nu}_e] = 
\f{ G_F^2 m_{b,pole}^5 \kappa(z) }{ 192 \pi^3 g(z) } |V_{cb}|^2 
\ee
where 
\be \label{g(z)}
g(z) = 1 - 8 z + 8 z^3 - z^4 - 12 z^2 \ln z \hspace{2cm} 
\left( z = \f{m_{c,pole}^2}{m_{b,pole}^2} \right)
\ee
is the phase-space factor, and 
\be  \label{kappa}
\kappa(z) = 1 - \f{2 \al (m_b)}{3 \pi} \f{h(z)}{g(z)}
\ee
is a sizable next-to-leading QCD correction to the semileptonic decay
\cite{CM78}. The function $h(z)$ has been given analytically in
ref.~\cite{N89}
\bea
h(z) =
- (1-z^2) \left( \f{25}{4}- \f{239}{3} z + \f{25}{4} z^2 \right) 
+ z \ln z \left( 20 + 90 z - \f{4}{3} z^2 + \f{17}{3} z^3 \right) 
+ z^2 \ln^2 z \; ( 36 + z^2) 
\hspace{1.5cm} && \nonumber \\
+ (1-z^2) \left( \f{17}{3} - \f{64}{3} z + \f{17}{3} z^2 \right) \ln (1-z) 
- 4 ( 1 + 30 z^2 + z^4 ) \ln z \; \ln (1-z) 
\hspace{5cm} && \nonumber \\
- (1 + 16z^2 + z^4) [ 6 {\rm Li}_2(z) - \pi^2 ] 
- 32 z^{3/2} (1+z) \left[ \pi^2 
     - 4 {\rm Li}_2(\sqrt{z}) +  4 {\rm Li}_2(-\sqrt{z}) 
     - 2 \ln z \; \ln \left( \f{1-\sqrt{z}}{1+\sqrt{z}} \right) \right]. 
\hspace{0.5cm} && \nonumber \eea 

	Thus, the final perturbative quantity we consider is the ratio
\be \label{ratio}
R_{quark}(\delta) = 
\f{ \Gamma[ b{\to}X_s\gamma]^{E_{\gamma} > (1-\delta) E_{\gamma}^{max}}}{
    \Gamma[ b \to X_c e \bar{\nu}_e ]} =
\f{|V_{ts}^* V_{tb}|^2}{|V_{cb}|^2} 
\f{6 \alpha_{em}}{\pi g(z)} F \left( |D|^2 + A \right).
\ee
where $D$ and $A$ are given in eqns. (\ref{Dvirt}) and (\ref{Abrem}),
respectively, and
\be \label{factor}
F = \f{1}{\kappa(z)} \left( \f{m_b(\mu=m_b)}{m_{b,pole}} \right)^2 = 
    \f{1}{\kappa(z)} \left( 1 - \f{8}{3} \f{\al(m_b)}{\pi} \right).
\ee

\vspace*{0.5cm}
\noindent {\bf 5.} Let us now turn to the numerical results. Besides
the five parameters listed above eqn.~(\ref{alpha.and.mt}), the
quantity $R_{quark}(\delta)$ in eqn.~(\ref{ratio}) depends on a few
more SM parameters. They are the following: \vspace{0.2cm}

\noindent (i) The ratio $m_{c,pole}/m_{b,pole}$. Similarly to
ref.~\cite{GHW96}, we will use
\be
\f{m_{c,pole}}{m_{b,pole}} = 0.29 \pm 0.02,
\ee
which is obtained from $m_{b,pole} = 4.8 \pm 0.15$~GeV and
$m_{b,pole}-m_{c,pole}=3.40$~GeV. \vspace{0.2cm}

\noindent (ii) The $b$-quark mass itself. Apart from the ratio
$m_c/m_b$, the $b$-quark mass enters our formulae only in two places:
in the explicit logarithm in $D$ (\ref{Dvirt}) and in the factor $F$
(\ref{factor}), as the argument of $\al$. In both cases, the
$m_b$-dependent terms are next-to-leading, and their $m_b$-dependence
is logarithmic. Changing $m_b$ from 4.6~GeV to 5~GeV in these places
affects \bsg decay rate by less than 2\%. Thus, it does not really
matter whether one uses the pole mass or the $\overline{MS}$ mass
there. We choose to use $m_{b,pole} = 4.8 \pm 0.15$~GeV.
\vspace{0.2cm}

\noindent (iii) The electromagnetic coupling constant $\alpha_{em}$.
It is not {\it a priori} known whether this constant should be
renormalized at $\mu \sim m_b$ or $\mu \sim M_W$, because the QED
corrections have not been included. We allow $\alpha_{em}$ to vary
between $\alpha_{em}(m_b)$ and $\alpha_{em}(M_W)$, i.e. we take
$\alpha_{em}^{-1} = 130.3 \pm 2.3$. \vspace{0.1cm}

\noindent (iv) The ratio of CKM factors. We will use
\be
\left| \f{V_{ts}^* V_{tb}}{V_{cb}} \right| = 0.976 \pm 0.010
\ee
given in ref. \cite{B96}. It corresponds to gaussian error analysis,
which is in accordance with interpreting the quoted error as a
$1\sigma$ uncertainty. \vspace*{0.1cm}

	The dependence of $R_{quark}(\delta)$ on $\delta$ is shown in
Fig.~1. The middle curve is the central value. The uncertainty of the
prediction is described by the shaded region. It has been found by
adding in squares all the parametric uncertainties mentioned above in
points (i)--(iv) and in eqns.~(\ref{alpha.and.mt})--(\ref{mu.range}).
\begin{figure}[h] 
\centerline{
\epsfysize = 10cm
\epsffile{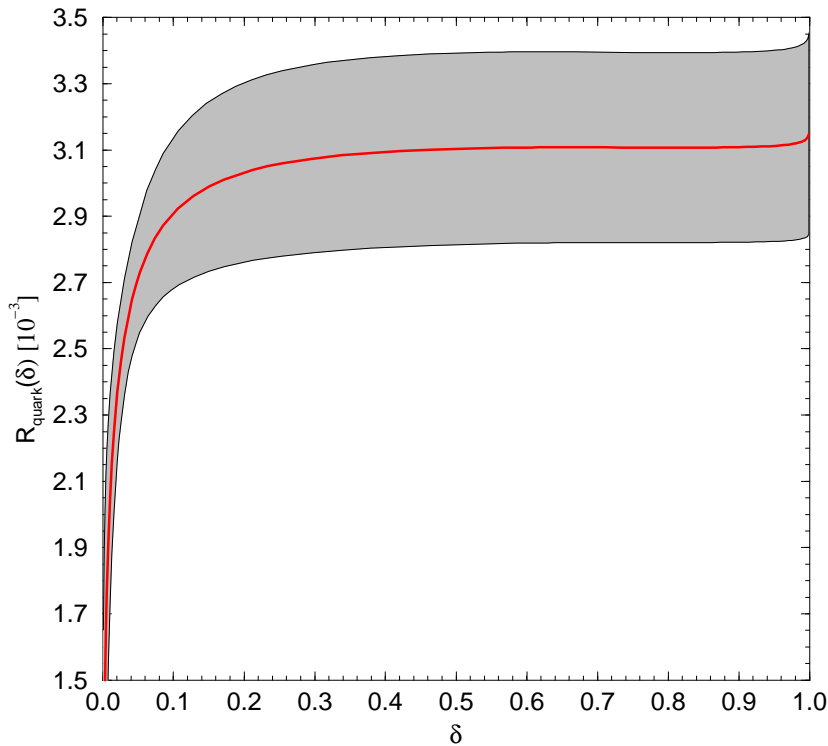}}
\caption{$R_{quark}(\delta)$ as a function of $\delta$.}
\end{figure}

	The ratio $R_{quark}(\delta)$ is divergent in the limit
$\delta{\to}1$, which is due to the function $f_{88}(\delta)$. In this
limit, the physical quantity to consider would be only the sum of \bsg
and $b{\to}X_s$ decay rates, in which this divergence would cancel out.

	The divergence at $\delta{\to}1$ is very slow. It manifests
itself only by a small kink in Fig.~1 where values of $\delta$ up to
0.999 were included. For $\delta = 0.99$, the contribution from
$f_{88}(\delta)$ to $R_{quark}(\delta)$ is well below 1\%. In order to
allow easy comparison with previous experimental and theoretical
publications, we choose $\delta^{max} = 0.99$ as the first particular
value of $\delta$ to consider. This value of $\delta$ will be assumed
in discussing the ``total'' $B{\to}X_s\gamma$ decay rate below.

	A value of $\delta$ which is closer to what is actually
measured is $\delta = z \equiv (m_c/m_b)^2$. It corresponds to
counting only the photons with energies above the charm production
threshold in the $b$-quark rest frame. This is the second value of
$\delta$ we will consider below.

	We find
\bea
R_{quark}(\delta^{max}) &=& ( 3.12 \pm 0.29 ) \times 10^{-3},\\ 	
R_{quark}(\delta = z)   &=& ( 2.86 \pm 0.24 ) \times 10^{-3}.	
\eea
In both cases, the uncertainty is below 10\%. The dominant sources of
it are $m_c/m_b$ and $\mu_b$. The relative importance of various
uncertainties is shown in Table.~1.
\begin{center}
\begin{tabular}{|l|c|c|c|c|c|c|c|}
\hline
& \multicolumn{7}{c|}{Source}\\
\cline{2-8}
& $\al(M_Z)$ & $m_{t,pole}$ & $\mu_b$ & $m_{c,pole}/m_{b,pole}$ & $m_{b,pole}$ & $\alpha_{em}$ & CKM angles \\
\hline
$\delta= \delta^{max}$ 
&    2.5\%   &    1.7\%   & 6.6\% &          5.2\%        &   0.5\%    &    1.9\%  & 2.1\%\\
$\delta=z$ 
&    2.2\%   &    1.7\%   & 3.3\% &          6.6\%        &   0.6\%    &    1.9\%  & 2.1\%\\
\hline
\end{tabular} 
\vspace{0.2cm}\\
{\small Table I. Uncertainties in $R_{quark}(\delta)$ due to various sources.}
\end{center}

	Let us note that setting $\delta = z$ introduces an additional
spurious inaccuracy due to $m_c/m_b$ in $R_{quark}$. However, this
additional uncertainty is significantly smaller than the original
inaccuracy due to $m_c/m_b$ for $\delta$ uncorrelated with $m_c/m_b$.

	It is quite surprising that the $\mu_b$-dependence is weaker
for $\delta = z$ than for $\delta = \delta^{max}$. Naively, one would
expect larger $\mu_b$-dependence for small values of $\delta$, when
the first term in eqn.~(\ref{Abrem}) becomes more and more important.
Clearly, $\delta = z$ is not small enough for this effect to show up.
The weaker $\mu_b$-dependence for $\delta = z$ seems to be quite
accidental.\\

\noindent {\bf 6.} In the end, we need to pass from the calculated
$b$-quark decay rates to the $B$-meson decay rates. Relying on the
Heavy Quark Effective Theory (HQET) calculations we write
\be \label{BR} 
{\cal B}(B{\to}X_s \gamma) = {\cal B}(B{\to}X_c e \bar{\nu_e})\;\;
R_{quark}(\delta^{max}) \;\; 
\left( 1 - \f{\delta^{NP}_{sl}}{m_b^2}
         + \f{\delta^{NP}_{rad}}{m_b^2} \right),
\ee
where $\delta^{NP}_{sl}$ and $\delta^{NP}_{rad}$
parametrize nonperturbative corrections to the semileptonic and
radiative $B$-meson decay rates, respectively. 

	Following refs.~\cite{FLS94,N95}, we express
$\delta^{NP}_{sl}$ and $\delta^{NP}_{rad}$ in terms of the HQET
parameters $\lambda_1$ and $\lambda_2$
\bea
\delta^{NP}_{sl}  &=& \f{1}{2} \lambda_1 + 
               \left( \f{3}{2} - \f{6 (1-z)^4}{g(z)} \right) \lambda_2,\\
\delta^{NP}_{rad} &=& \f{1}{2} \lambda_1 - \f{9}{2} \lambda_2,
\eea
where $g(z)$ is the same as in eqn. (\ref{g(z)}). The poorly known
parameter $\lambda_1$ cancels out in the r.h.s. of eqn. (\ref{BR})
which depends only on the difference $\delta^{NP}_{sl} -
\delta^{NP}_{rad}$. The value of $\lambda_2$ is known from $B^*$--$B$
mass splitting
\be
\lambda_2 = \f{1}{4} ( m_{B^*}^2 - m_B^2 ) \simeq 0.12\;{\rm GeV}^2.
\ee

	The two nonperturbative corrections in eqn.~(\ref{BR}) are
both around $4\%$ in magnitude, and they tend to cancel each other. In
effect, they sum up to only around $ 1 \%$. Such a small number has to
be taken with caution, because the four-quark operators $P_1$, ...,
$P_6$ have not been included in the calculation of
$\delta^{NP}_{rad}$. Contributions from these operators could
potentially give one- or two-percent effects. Nevertheless, it seems
reasonable to conclude that the total nonperturbative correction to
eqn.~(\ref{BR}) is well below 10\%, i.e. it is smaller than the
inaccuracy of the perturbative calculation of
$R_{quark}(\delta^{max})$.

	When the above caution is ignored and ${\cal B}(B{\to}X_c e
\bar{\nu_e}) = ( 10.4 \pm 0.4)\%$~\cite{PDG96} is used, one obtains
the following numerical prediction for $B{\to}X_s \gamma$ branching
ratio
\be
{\cal B}(B{\to}X_s \gamma) = (3.28 \pm 0.33) \times 10^{-4}.
\ee

	The central value of this prediction is outside the $1\sigma$
experimental error bar in eqn.~(\ref{CLEO}).\footnote{We identify the
$b$-quark decay rate given by CLEO with the $B$-meson decay rate.
Large statistical errors in the CLEO result make this identification
acceptable.} However, the experimental and theoretical error bars
practically touch each other. Therefore, we conclude that the present
$B{\to}X_s\gamma$ measurement remains in agreement with the Standard
Model. This conclusion holds in spite of that the theoretical
uncertainty is now more than twice smaller than in the previously
available leading order prediction~\cite{BMMP94}.

	In the measurement of $B{\to}X_s\gamma$ branching fraction,
one needs to choose a certain lower bound on photon energies. Instead
of talking about the ``total'' decay rate, it is convenient to count
only the photons with energies above the charm production threshold in
the $B$-meson rest frame
\be \label{threshold}
E_{\gamma} > \f{m_B^2 - m_D^2}{2 m_B} \simeq 2.31\;{\rm GeV}.
\ee
Most of the photons in $B{\to}X_s\gamma$ survive this energy cut, and
a huge background from charm production is removed. Let us denote the
corresponding branching fraction by ${\cal B}(B{\to}X_s
\gamma)^{above}$.

	If the $b$-quark was infinitely heavy, it would not move
inside the $B$-meson. The restriction (\ref{threshold}) would then be
equivalent to setting $\delta = z$ at the quark level. Since the Fermi
motion of the $b$-quark inside the $B$-meson is a ${\cal
O}(\bar{\Lambda}/m_b)$ effect, we can write similarly to
eqn.~(\ref{BR})
\be \label{BRabove} 
{\cal B}(B{\to}X_s \gamma)^{above} = {\cal B}(B{\to}X_c e \bar{\nu})\;\;
R_{quark}(\delta=z) \;\; ( 1 + {\cal O}(\bar{\Lambda}/m_b)).
\ee

	The ${\cal O}(\bar{\Lambda}/m_b)$ part of the latter equation
can be calculated using specific models for the Fermi motion of the
$b$-quark inside the $B$-meson, as it has been done e.g. in
ref.~\cite{AG95} (see also ref. \cite{N94}). Performing a similar
calculation with use of the NLO values of the Wilson coefficients is
beyond the scope of the present paper. Such an analysis will be
necessary when more statistics allows to reduce experimental errors in
measurements of ${\cal B}(B{\to}X_s\gamma)^{above}$.\\

\noindent {\bf 7.} To conclude, we have presented practically complete
NLO formulae for \bsg decay in the Standard Model. They include
previously published contributions as well as our new results for
three-loop anomalous dimensions. Our prediction for $B{\to}X_s\gamma$
branching fraction in the Standard Model is $(3.28 \pm 0.33) \times
10^{-4}$ which remains in agreement with the CLEO measurement at the
$2 \sigma$ level. Clearly, an interesting test of the SM will be
provided when more precise experimental data are available. More
importantly, the $B{\to}X_s\gamma$ mode will have more exclusion power
for extensions of the SM.

\newpage
\noindent {\bf \Large Appendix A.}

	Here, we give the eight evanescent operators we have used in
our anomalous dimension computation. Giving them explicitly is
necessary in order to fully specify our renormalization scheme, and
thus give meaning to the anomalous dimension matrix presented in
eqn.~(\ref{gamma1}).
\bea 
P_{11} &=& (\bar{s}_L \gamma_{\mu_1}
                      \gamma_{\mu_2}
                      \gamma_{\mu_3} T^a c_L)(\bar{c}_L \gamma^{\mu_1} 
                                                        \gamma^{\mu_2}
                                                        \gamma^{\mu_3} T^a b_L) 
-16 P_1 \nonumber \\
P_{12} &=& (\bar{s}_L \gamma_{\mu_1}
                      \gamma_{\mu_2}
                      \gamma_{\mu_3}     c_L)(\bar{c}_L \gamma^{\mu_1} 
                                                        \gamma^{\mu_2}
                                                        \gamma^{\mu_3}     b_L) 
-16 P_2 \nonumber \\
P_{15} &=& (\bar{s}_L \gamma_{\mu_1}
                      \gamma_{\mu_2}
                      \gamma_{\mu_3}
                      \gamma_{\mu_4}
                      \gamma_{\mu_5}     b_L)\sum_q(\bar{q} \gamma^{\mu_1} 
                                                            \gamma^{\mu_2}
                                                            \gamma^{\mu_3}
                                                            \gamma^{\mu_4}
                                                            \gamma^{\mu_5}     q) 
-20 P_5 + 64 P_3 \nonumber \\
P_{16} &=& (\bar{s}_L \gamma_{\mu_1}
                      \gamma_{\mu_2}
                      \gamma_{\mu_3}
                      \gamma_{\mu_4}
                      \gamma_{\mu_5} T^a b_L)\sum_q(\bar{q} \gamma^{\mu_1} 
                                                            \gamma^{\mu_2}
                                                            \gamma^{\mu_3}
                                                            \gamma^{\mu_4}
                                                            \gamma^{\mu_5} T^a q) 
-20 P_6 + 64 P_4 \nonumber \\
P_{21} &=& (\bar{s}_L \gamma_{\mu_1}
                      \gamma_{\mu_2}
                      \gamma_{\mu_3}
                      \gamma_{\mu_4}
                      \gamma_{\mu_5} T^a c_L)(\bar{c}_L \gamma^{\mu_1} 
                                                        \gamma^{\mu_2}
                                                        \gamma^{\mu_3}
                                                        \gamma^{\mu_4}
                                                        \gamma^{\mu_5} T^a b_L) 
-20 P_{11} - 256 P_1 \\
P_{22} &=& (\bar{s}_L \gamma_{\mu_1}
                      \gamma_{\mu_2}
                      \gamma_{\mu_3}
                      \gamma_{\mu_4}
                      \gamma_{\mu_5} T   c_L)(\bar{c}_L \gamma^{\mu_1} 
                                                        \gamma^{\mu_2}
                                                        \gamma^{\mu_3}
                                                        \gamma^{\mu_4}
                                                        \gamma^{\mu_5}     b_L) 
-20 P_{12} - 256 P_2 \nonumber \\
P_{25} &=& (\bar{s}_L \gamma_{\mu_1}
                      \gamma_{\mu_2}
                      \gamma_{\mu_3}
                      \gamma_{\mu_4}
                      \gamma_{\mu_5}
                      \gamma_{\mu_6}
                      \gamma_{\mu_7}     b_L)\sum_q(\bar{q} \gamma^{\mu_1} 
                                                            \gamma^{\mu_2}
                                                            \gamma^{\mu_3}
                                                            \gamma^{\mu_4}
                                                            \gamma^{\mu_5}
                                                            \gamma^{\mu_6}
                                                            \gamma^{\mu_7}     q) 
-336 P_5 + 1280 P_3 \nonumber \\
P_{26} &=& (\bar{s}_L \gamma_{\mu_1}
                      \gamma_{\mu_2}
                      \gamma_{\mu_3}
                      \gamma_{\mu_4}
                      \gamma_{\mu_5}
                      \gamma_{\mu_6}
                      \gamma_{\mu_7} T^a b_L)\sum_q(\bar{q} \gamma^{\mu_1} 
                                                            \gamma^{\mu_2}
                                                            \gamma^{\mu_3}
                                                            \gamma^{\mu_4}
                                                            \gamma^{\mu_5}
                                                            \gamma^{\mu_6}
                                                            \gamma^{\mu_7} T^a q) 
-336 P_6 + 1280 P_4. \nonumber 
\eea

\vspace{1cm}
\noindent {\bf \Large Appendix B.}

	Here, we express the functions $f_{ij}(\delta)$ present in
eqn.~(\ref{Abrem}) in terms of the quantities $M_{kl}(t,u)$ given in
eqn.~(27) of ref.~\cite{P96}.\footnote{
There is a global factor of 2 missing in the expression for $M_{78}$
given in eqn.~(27) of ref.~\cite{P96}.}
For each $(ij) \neq (77)$ and $i \leq j$, $f_{ij}(\delta)$ is given by
\be
f_{ij}(\delta) = \f{2}{3} (2-\delta_{ij}) 
\sum_{ \stackrel{k,l=1}{k \leq l}}^8 W_{ik} 
\left[ \int_{1-\delta}^1 du \int_{1-u}^1 dt\; M_{kl}(t,u) \right] W_{jl},
\ee
where 
\be
\hat{W} = \left[ \begin{array}{cccccccc} 
\f{1}{2} & -\f{1}{6} &      0    &     0    &      0    &     0    & 0 & 0\\
    0    &      1    &      0    &     0    &      0    &     0    & 0 & 0\\
    0    &      0    &      1    &     0    &      1    &     0    & 0 & 0\\
    0    &      0    & -\f{1}{6} & \f{1}{2} & -\f{1}{6} & \f{1}{2} & 0 & 0\\
    0    &      0    &     16    &     0    &      4    &     0    & 0 & 0\\
    0    &      0    & -\f{8}{3} &     8    & -\f{2}{3} &     2    & 0 & 0\\
    0    &      0    &      0    &     0    &      0    &     0    & 1 & 0\\
    0    &      0    &      0    &     0    &      0    &     0    & 0 & 1
\end{array} \right] \ee
In four space-time dimensions, the matrix $\hat{W}$ transforms our
operator basis (\ref{ope}) into the one used in ref.~\cite{P96}
\be
P_i \;\; \stackrel{D=4}{= \hspace{-0.1cm} = \hspace{-0.1cm} =} 
\;\; \sum_{k=1}^8 W_{ik} \hat{O}_k.
\ee

\newpage
\setlength {\baselineskip}{0.2in}
 
\end{document}